\documentclass[12pt]{iopart}

\usepackage{iopams} 
\usepackage{graphicx}

\begin{document}

\title[Methods for measuring the electron EDM using ultracold YbF molecules]{Methods for measuring the electron's electric dipole moment using ultracold YbF molecules}
\author{N. J. Fitch\footnote{These authors contributed equally to this work.}\addtocounter{footnote}{-1}, J. Lim\footnotemark, E. A. Hinds, B. E. Sauer and M. R. Tarbutt}
\address{Centre for Cold Matter, Blackett Laboratory, Imperial College London, Prince Consort Road, London SW7 2AZ, United Kingdom}
\eads{\mailto{m.tarbutt@imperial.ac.uk}, \mailto{n.fitch@imperial.ac.uk}}

\begin{abstract}
Measurements of the electron's electric dipole moment (eEDM) are demanding tests of physics beyond the Standard Model. We describe how ultracold YbF molecules could be used to improve the precision of eEDM measurements by two to three orders of magnitude. Using numerical simulations, we show how the combination of magnetic focussing, two-dimensional transverse laser cooling, and frequency-chirped laser slowing, can produce an intense, slow, highly-collimated molecular beam. We show how to make a magneto-optical trap of YbF molecules and how the molecules could be loaded into an optical lattice. eEDM measurements could be made using the slow molecular beam or using molecules trapped in the lattice. We estimate the statistical sensitivity that could be reached in each case and consider how sources of noise can be reduced so that the shot-noise limit of sensitivity can be reached. We also consider systematic effects due to magnetic fields and vector light shifts and how they could be controlled.
\end{abstract}

\vspace{2pc}
%
%
%
%

\section{Introduction}

Cosmological observations have established that the visible Universe is almost entirely devoid of antimatter. This asymmetry between matter and antimatter cannot be understood in the framework of the known fundamental forces, since they treat matter and antimatter on a very nearly equal footing. In the context of the three discrete symmetries -- charge conjugation (C), parity (P) and time-reversal (T) -- the formation of an excess of matter requires CP violation~\cite{Sakharov1991}. The CP-violating interactions in the Standard Model of particle physics fail, by many orders of magnitude, to explain the excess of matter in the Universe. As a result, it is widely accepted that there must be undiscovered physics that introduces new CP-violating interactions, characterized by new particles of energy $\Lambda$ and CP-violating phases $\phi_{\rm CP}$. Many theories that extend the Standard Model to address this, including most supersymmetric models, give the electron an electric dipole moment (eEDM, $d_{e}$) at the one-loop level: $d_{e} \sim (10^{-26}\,e\rm{~cm})\,\sin \phi_{\rm CP}/\Lambda^{2}$, with $\Lambda$ in TeV. Such models normally expect $\Lambda$ to be greater than a few hundred GeV, while $\phi_{\rm CP}$ has no reason to be particularly small, so these new forces could be revealed by precise measurements of $d_{e}$. By contrast, the Standard Model predicts an exceedingly tiny value for the eEDM, $d_{e} < 10^{-38} \,e$~cm~\cite{Pospelov2014}, which is too small to measure. 

Measurements of the eEDM have steadily improved in sensitivity over many years, but so far all have yielded values consistent with zero. Using ultracold polar molecules, it seems feasible to measure $d_{e}$ with a precision of $10^{-32}\,e$~cm. Such a measurement would be a search for the CP-violating interactions responsible for the matter-antimatter asymmetry, and a test of physics beyond the Standard Model up to $\Lambda \sim 1000$~TeV. This energy lies beyond the kinematic reach of any current or projected particle collider, and is beyond the favoured mass range for the super-partners of quarks and leptons.

The electron's electric dipole moment can be detected through the precession of the electron spin in an applied electric field. Because the precession rate is greatly enhanced for electrons in some heavy atoms~\cite{Hinds1997}, all the early measurements used such systems~\cite{Regan2002}. Still higher enhancement is available in heavy polar molecules, and the most precise eEDM measurements have all been made this way. Measurements have been completed using beams of YbF and ThO molecules~\cite{Hudson2011, Baron2014, Andreev2018}, a cell of hot PbO vapour~\cite{Eckel2013}, and trapped HfF$^{+}$ molecular ions~\cite{Cairncross2017}. In 2018, the ACME collaboration used a beam of ThO molecules to set the current best upper limit of $|d_e| < 1.1 \times 10^{-29} \,e$~cm~\cite{Andreev2018}. It is remarkable that the eEDM should still be zero at this tiny level. Many theories of the new forces, such as the simpler forms of supersymmetry, predict considerably larger values of $d_e$, so appear to be ruled out by this result.

The statistical uncertainty in an ideal shot-noise limited measurement of $d_e$, in units of $e$\,cm, is
\begin{equation}
\sigma_{d_e} = \frac{\hbar}{2 e E_{\rm eff}\tau \sqrt{N_{\rm det}}},
\label{eq:statSensitivity}
\end{equation}
where $E_{\rm eff}$ is the effective electric field (in V/cm), $\tau$ is the spin-precession time, and $N_{\rm det}$ is the number of particles detected.  The measurement uncertainty is minimized by detecting as many molecules as possible and using the longest possible spin-precession time.  The ThO experiment uses a molecular beam with a high flux of metastable molecules, but the spin precession time is limited by the 1.8~ms natural lifetime of the metastable state~\cite{Vutha2010}. The HfF$^+$ experiment uses trapped molecular ions with spin precession times close to 1~s, but the Coulomb interaction limits the number of ions that can be used. By contrast, it seems possible  to obtain both large $\tau$ and large $N_{\rm det}$ using neutral molecules cooled to a temperature of a few $\mu$K.  Recently, several groups have shown how to apply laser cooling directly to molecules~\cite{Shuman2010, Hummon2013, Zhelyazkova2014, Barry2014, Truppe2017b, Anderegg2017}, reaching temperatures as low as 5~$\mu$K~\cite{Cheuk2018, Caldwell2019}. Importantly, the method can be applied to the polar, paramagnetic molecules needed for an eEDM measurement. Electron EDM experiments using ultracold YbF~\cite{Tarbutt2013}, YbOH~\cite{Kozyryev2017b}, and BaF~\cite{Aggarwal2018} are all currently being developed, and a nuclear EDM experiment using TlF molecules is also underway~\cite{Norrgard2017}. Laser cooling of YbF~\cite{Lim2018} and YbOH~\cite{Augenbraun2020} have already been demonstrated.

In this paper, we explore how laser-cooled YbF molecules could be used to make a sensitive measurement of $d_e$.  We consider two approaches, both currently being pursued in our laboratory.  In the first, the measurement is carried out using a slow molecular beam that is collimated by transverse laser cooling.  In the second approach, molecules are captured in a magneto-optical trap (MOT), cooled to $\mu$K temperature, and loaded into an optical lattice where the measurement takes place.  We consider how to maximise the number of molecules in each of these experiments, and model each step in order to provide realistic sensitivity estimates. We compare the relative strengths and challenges of the two approaches. We also consider some  challenging systematic effects and how they could be controlled.  

\section{Cryogenic source of molecules}
\label{sec:source}
Currently, all experiments based on laser-cooled molecules begin with a cryogenic buffer-gas beam~\cite{Hutzler2012}.  The high flux, low temperature, and relatively low forward speed of these beams are all relevant for producing a large number of ultra-cold molecules.  Our current source, based on a single-stage cell design, produces a beam with a speed of 160~m/s and a flux of 2$\times$10$^{10}$ molecules per steradian per pulse in a single internal state~\cite{Truppe2017c}. The YbF molecules are produced by laser ablation of Yb in the presence of SF$_6$ gas, and are then swept out of the cell by a flow of helium buffer gas that is cooled to 4~K. Several improvements can be implemented to lower the speed and improve the flux. First, the cell temperature can be reduced to about 1.5~K by using the latest developments in closed-cycle cryocooler technology. This should lower the speed of the beam as well as the initial internal rovibrational temperature. Recently, a cryogenic source cooled to 2~K produced $10^9$ YbOH molecules per pulse in a single rotational state, with a mean forward speed of 90~m/s and a transverse velocity spread of 15~m/s, corresponding to a flux of about $10^{10}$ molecules per steradian per pulse~\cite{Augenbraun2020}. Second, a two-stage cell design can be used to lower the speed even further~\cite{Hutzler2012}. Beam speeds as low as 40~m/s have been reached in this way, although with a significant reduction in flux. Third, it seems likely that the flux can be increased by enhancing the chemical reaction between Yb and SF$_6$ using laser light resonant with the $^1$S$_0\leftrightarrow^3$P$_1$ Yb transition at 556 nm. This method was recently demonstrated for the chemically similar molecule YbOH, where a factor of 10 increase in flux was obtained~\cite{Jadbabaie2020}. This scheme may also reduce the forward speed of the beam, since high speed collisions will no longer be needed to produce the molecules efficiently.  Implementing all of these advances should yield a YbF molecular beam with a forward speed of about 80~m/s and a flux of 10$^{10}$--10$^{11}$ molecules per steradian per pulse in the desired quantum state.

\section{Laser cooling scheme}

\begin{figure}[t]
	\centering
	\includegraphics[width=\columnwidth]{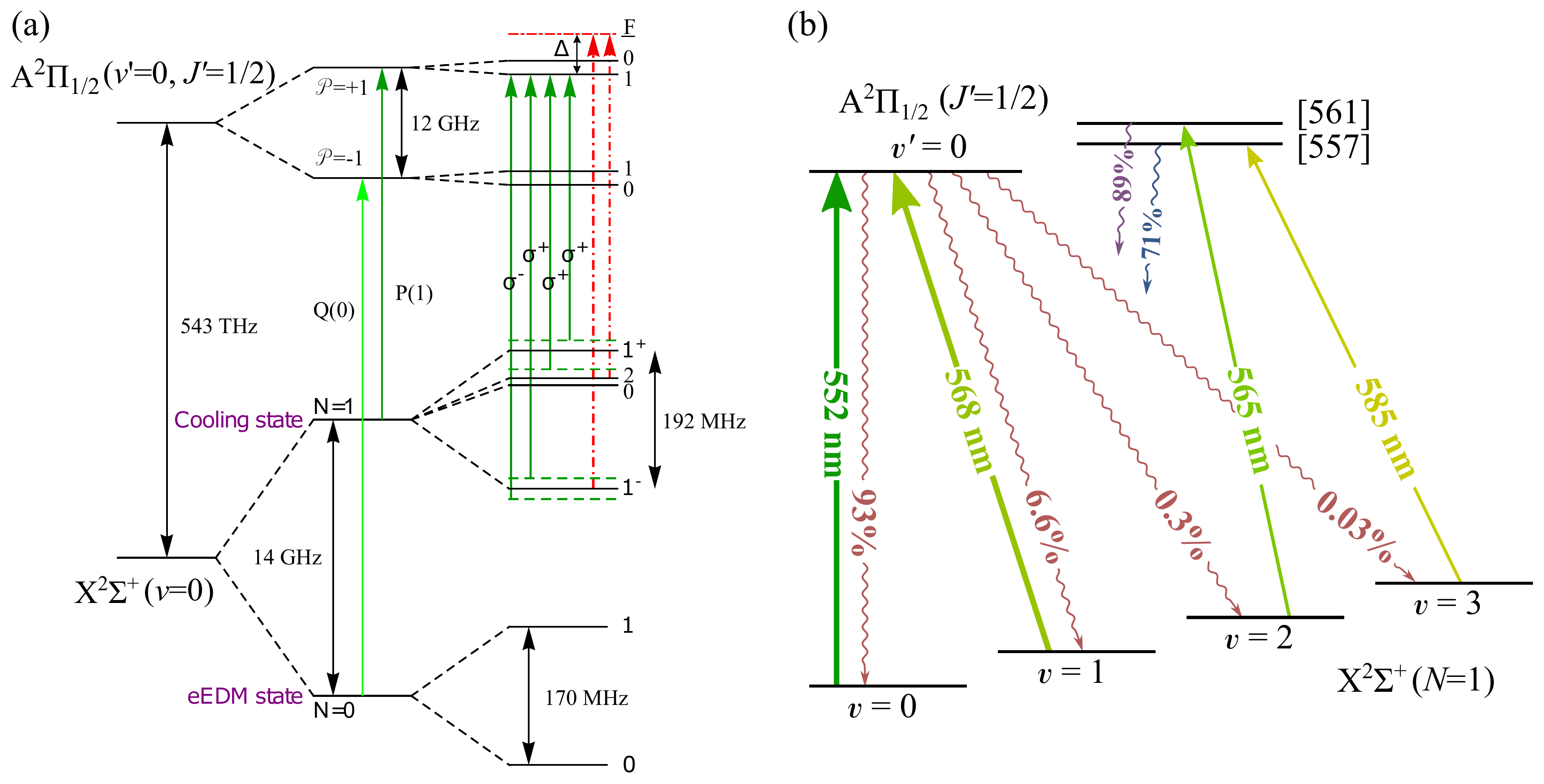}
	\caption{(a) Energy levels of YbF most relevant to the experiment. The quantum numbers $v$, $N$, $J$, $F$ and ${\cal P}$ are  for vibration, rotation, total electronic angular momentum, total angular momentum, and parity. The eEDM experiment uses the hyperfine components of the lowest rotational state, $N=0$. Laser cooling uses the P(1) transition from $N=1$. The hyperfine components of this transition are labelled with $\sigma^{\pm}$ to indicate the polarizations needed for the magneto-optical trap discussed in Sec.~\ref{Sec:MOT}. Similarly, the red dash-dotted lines indicate the sideband structure used for the transverse laser cooling scheme discussed in Sec.~\ref{Sec:Transverse}. (b) Energy levels and branching ratios for the laser cooling of YbF. Solid arrows indicate transitions used for laser cooling and repumping, along with their wavelengths. Wavy arrows indicate spontaneous decay pathways with their branching ratios. \label{Fig:EnergyLevels}}
\end{figure}

Figure~\ref{Fig:EnergyLevels}(a) shows the energy levels of the YbF molecule that are most relevant to our discussion. The eEDM measurement mainly uses the lowest-lying state of the molecule, X$^{2}\Sigma^{+}(v=0,N=0)$. The main laser cooling transition is from the first rotationally excited state of the ground electronic state, labelled $X^{2}\Sigma^{+} (v=0,N=1)$, to the lowest positive parity level of the excited electronic state, labelled  $A^{2}\Pi_{1/2} (v'=0,J'=1/2)$. This transition, labelled P(1) in the figure, has a natural linewidth of 5.7~MHz. The selection rules for angular momentum and parity ensure that the excited state can only decay back to the $N=1$ level of X, so the transition is rotationally closed. The hyperfine components of this transition are also shown. The dark green arrows indicate the detunings and polarizations of the frequency components needed for magneto-optical trapping (see section \ref{Sec:MOT}), and the red dash-dotted arrows indicate the frequnecy components used for transverse cooling (section \ref{Sec:Transverse}).

Figure~\ref{Fig:EnergyLevels}(b) illustrates the laser cooling scheme in more detail. Although the cooling transition is rotationally closed, it is not vibrationally closed -- the excited state can decay to higher-lying vibrational states of X, albeit with rapidly diminishing probability as $v$ increases. To keep molecules in the cooling cycle, repump lasers are used to drive the equivalent rotationally closed transitions from $v=1,2$ and 3. The lowest vibrational level of the A state is a pure $^2\Pi_{1/2}$ state, but the higher vibrational states are mixtures of two electronic configurations, one involving excitation of the 6s electron and the other excitation of the 4f electron~\cite{Sauer1999, Lim2017}. The first two such states are labelled [557] and [561] in figure~\ref{Fig:EnergyLevels}(b). In our laser cooling scheme, we use these mixed levels for repumping the $v=2$ and $v=3$ states. The main cooling transition and the repump transitions each have four hyperfine components, with energy splittings that depend on $v$. In order to address these, radio-frequency sidebands are added to each laser using a combination of acousto-optic and electro-optic modulators customized for each transition. The complete laser and optical system is described in detail in~\cite{AlmondThesis}.

\section{Transverse laser cooling\label{Sec:Transverse}}

Using the cooling scheme described above, we have previously demonstrated laser cooling of a YbF beam in one transverse direction~\cite{Lim2018}. The molecules were cooled by sub-Doppler processes and we determined an upper temperature limit of 100~$\mu$K. These results show that YbF molecules can be cooled to the ultracold temperatures needed for the eEDM experiments discussed in this manuscript.

\begin{figure}[t]
	\centering
	\includegraphics[width=\columnwidth]{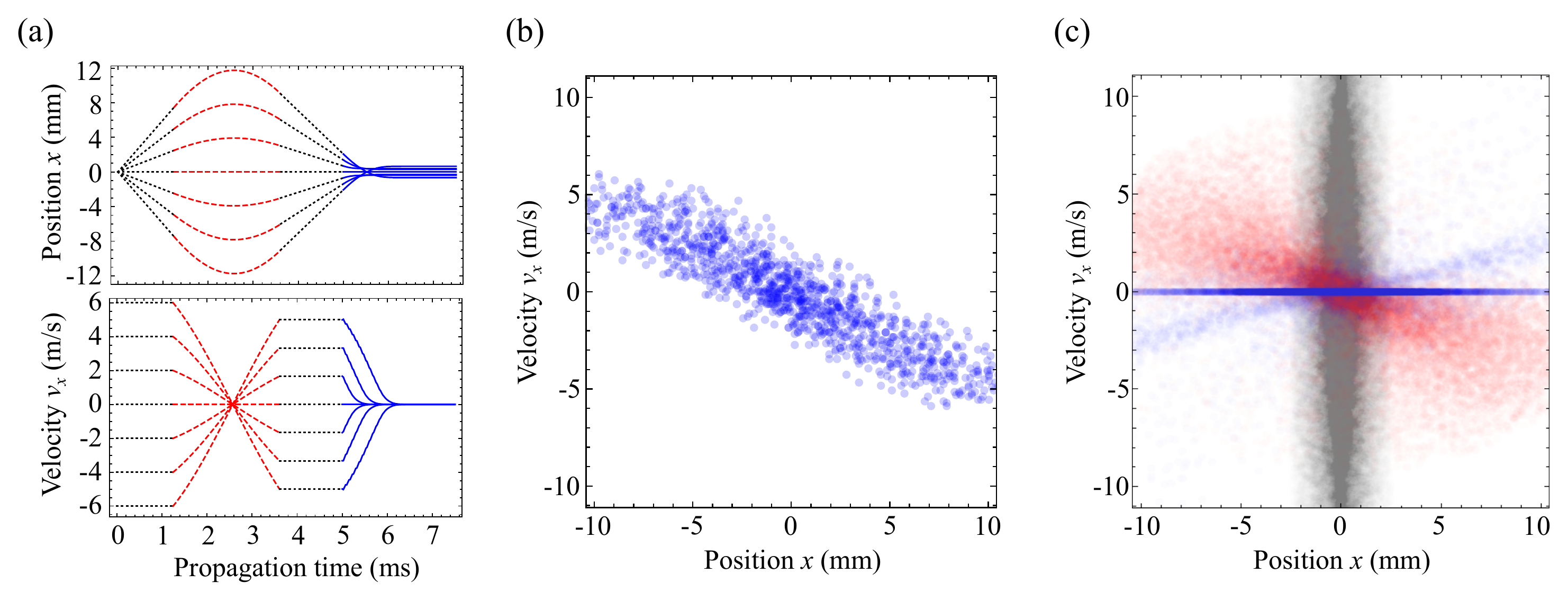}
	\caption{Numerical simulations of magnetic focussing followed by 2D transverse laser cooling. (a) Schematic representation of trajectories with various initial transverse velocities. There are three sections:  free flight (grey dotted lines), magnetic focussing (red dashed lines), and transverse laser cooling (blue solid lines). (b) Phase space acceptance of a 20-cm-long, 2D, transverse cooling region. The dots show the initial phase space distribution of those molecules that are finally cooled to ultracold temperature. (c) Phase space distributions at various points. Grey: output from the buffer gas source. Red: output from the magnetic lens designed to match the phase space to the acceptance plotted in (b). Blue: after the laser cooling.\label{Fig:focusAndCool}}
\end{figure}

Here, we consider how to maximize the number of ultracold molecules in the beam by extending the laser cooling to two dimensions and combining it with magnetic focussing. The idea is illustrated in figure~\ref{Fig:focusAndCool}(a). A magnetic lens is used to capture a large fraction of the molecules emitted from the cryogenic source and focus them into the laser cooling region. Here, 2D laser cooling acts as a collimator, resulting in a beam with small transverse spread in both position and velocity. We test these ideas using Monte-Carlo trajectory simulations of the focussing and transverse cooling. A set of permanent magnets placed 100~mm from the cell aperture produces a magnetic lens with a harmonic potential inside a cylinder of length 189~mm and radius 14~mm. The field at the walls is 1.5~T, which corresponds to a transverse capture velocity of $\sim$9.3~m/s for molecules in weak-field-seeking quantum states.  Half the molecules emitted from the source seek a weak field, and the other half could be captured by employing optical pumping before the lens. After exiting the focussing magnets, the molecules propagate freely for 111~mm before reaching the laser-cooling region.  Here, the 2D laser cooling force is calculated as a function of velocity and laser intensity by solving the multi-level optical Bloch equations, extending the approach of Refs.~\cite{Lim2018} and~\cite{Devlin2016}. The main cooling light has two frequency components separated by 159~MHz and detuned from the $F=2$ and $F=1^{-}$ states by 34~MHz, as indicated by the red arrows in figure \ref{Fig:EnergyLevels}(a). The laser light propagating on each transverse axis is polarised with the linear-parallel-linear configuration discussed in~\cite{Lim2018}. The Gaussian laser cooling beams have a $1/e^2$ intensity diameter of 5.6~mm, and the total power of the main cooling light is 400~mW shared equally between the two sidebands. A magnetic field of 5.0~G is applied, oriented at 45 degrees with respect to the laser polarization axis and the molecular beam axis, throughout the laser cooling region. 

The points in figure~\ref{Fig:focusAndCool}(b) show the phase-space acceptance of the simulated transverse laser cooling stage. To produce this plot, we simulate the trajectories of molecules through the 20-cm-long laser cooling region, and select those whose final transverse speeds and positions are below 1~cm/s and $\pm 5$~mm, respectively. In the simulation, molecules arriving at the laser-cooling region with transverse speeds below 6~m/s and initial positions between $\pm 10$~mm can be cooled into the ultracold regime. Those that can be cooled have a negative correlation between position and velocity, corresponding to a converging beam. Figure~\ref{Fig:focusAndCool}(c) shows the phase space distributions at various stages of the cooling process. The output from the buffer gas source (grey  points) is assumed to have a Gaussian forward velocity distribution centered at 80~m/s with a full width at half maximum (FWHM) of 42~m/s. Initial transverse positions are chosen from a Gaussian distribution of 4.2~mm FWHM, corresponding to the 5~mm diameter exit aperture of the cell, with associated velocities randomly sampled assuming a 1.8~K transverse temperature.  This initial distribution is transformed by the magnetic lens into the one shown by the red points, which closely matches the phase-space acceptance of the laser cooling region. The molecules are then laser-cooled to produce a beam roughly 4.1~mm in diameter (FWHM), with transverse velocities of $\sim 1$~cm/s (blue points). The blue points also show a weak diverging beam. This consists of molecules that were outside the acceptance of the laser cooling stage. With this apparatus, the simulation predicts that more than 12\% of all molecules leaving the source can be cooled to ultracold transverse temperature. Without the magnetic lens only 0.6\% would be cooled. The combination of magnetic focussing followed by 2D transverse cooling has not been described before, and we see that it is a powerful way to produce a high flux of ultracold molecules. Assuming an initial flux of 10$^{10}$ molecules per steradian per pulse in the $N=1$ state, which is at the lower end of the range estimated in section~\ref{sec:source}, there will be $\sim$10$^{8}$ ultracold molecules in each pulse of the beam.

\section{Laser slowing \label{Sec:Slowing}}

\begin{figure}[t]
	\centering
	\includegraphics[width=\columnwidth]{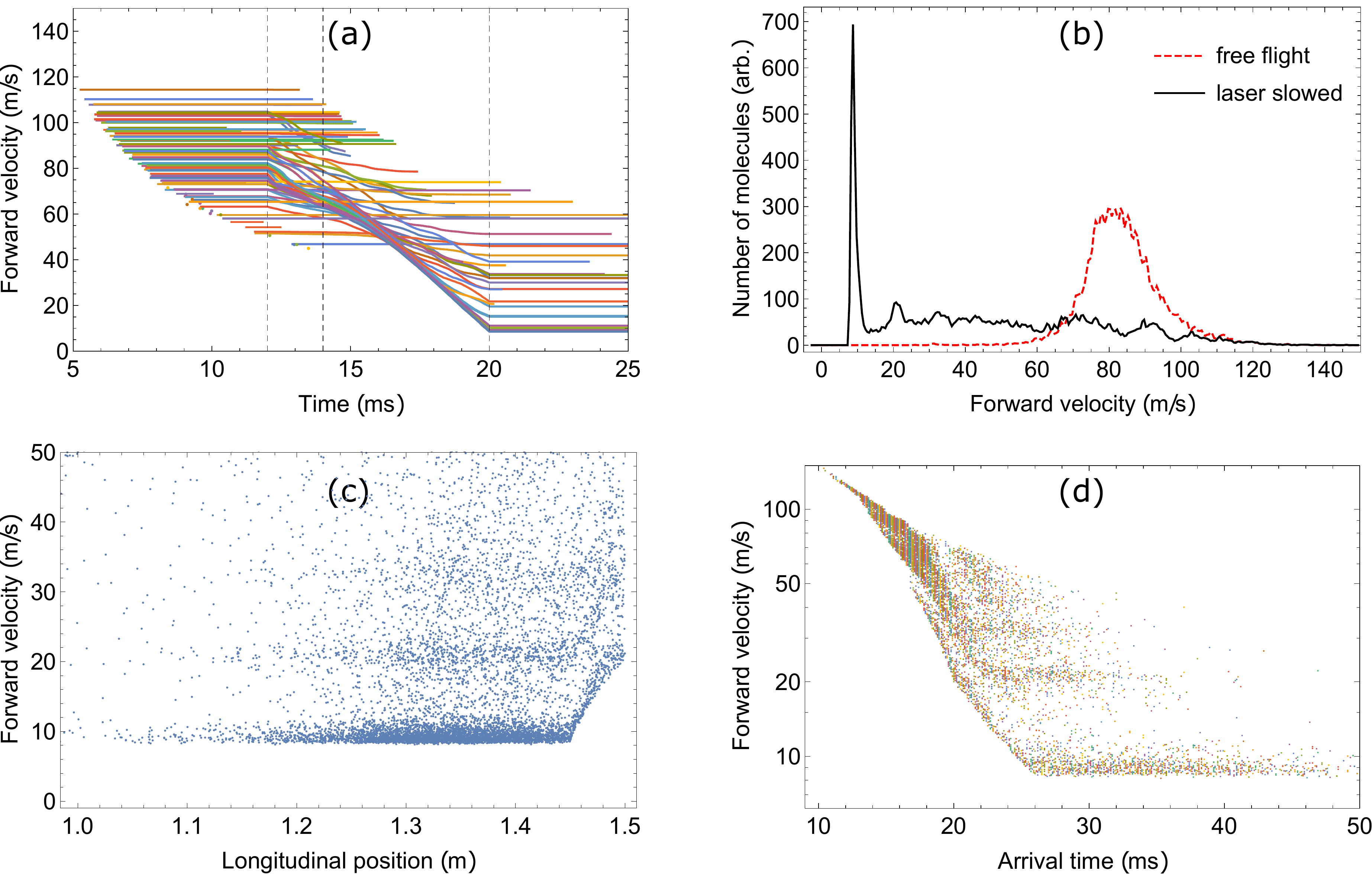}
	\caption{Simulations of laser slowing of YbF molecules using the frequency-chirped method. (a)  Trajectories of individual molecules during slowing. Dashed lines indicate the times when the slowing laser is turned on, begins the chirp, and turns off. (b) Simulated velocity distributions of molecules arriving 1.4~m from the buffer-gas cell, both with (black curve) and without (red dashed curve) laser slowing applied. (c) Longitudinal phase-space distribution of molecules after laser-slowing is complete. (d) Arrival time and forward velocity of molecules that arrive at the MOT position.}
	\label{Fig:LaserSlowingFigure}
\end{figure}

We have shown above that magnetic focusing followed by transverse laser cooling will produce a 4~mm-diameter, highly collimated molecular beam. Such a beam is eminently suitable for slowing by laser light propagating in the opposite direction. As the molecules slow down, we can deal with the changing Doppler shift either by chirping the laser frequency or by using frequency-broadened light. We have simulated the laser slowing of the molecules emerging from the transverse laser cooling region. In these simulations, the frequency-chirped method is employed to slow the beam to below 10~m/s.  The simulation takes into account the full hyperfine and Zeeman structure of the X and A states. The transitions are driven by a laser with 250~mW of power equally spread among 12 radio-frequency sidebands spaced by 20~MHz.  Leaks from the cooling/slowing cycle are neglected. The initial phase-space coordinates of the molecules are taken from the outputs of the transverse-cooling simulations. This modelling gives a scattering rate of $3\times 10^6$ photons/s, which produces an acceleration of $-11$~km/s$^2$. Molecules with an initial speed of 80~m/s will be brought to rest in 7.3~ms, after travelling a distance of 30~cm and scattering around $2 \times 10^4$~photons.

Figure~\ref{Fig:LaserSlowingFigure}(a) illustrates how the forward velocities of a set of molecules change over time. Here, the slowing laser is turned on 12~ms after the start of the beam pulse, which is late enough that nearly all molecules have already passed through the transverse laser-cooling region. The slowing sequence consists of holding the laser frequency fixed for the first 2~ms in order to reduce the forward velocity spread, then chirping the frequency at 20~MHz/ms for 6~ms to track the changing Doppler shift, at which point the laser is turned off.  The timings of the slowing sequence are shown as vertical dashed lines in the figure. We see that molecules with initial speeds between 40 and 100~m/s are brought into the desired velocity range at nearly the same time. Figure~\ref{Fig:LaserSlowingFigure}(b) shows the velocity distribution of molecules arriving at a detector that is 1.4~m from the buffer gas source. Only molecules passing within a 5~mm diameter circle centred on the longitudinal axis are counted. The red dashed curve shows the distribution without laser slowing. This is slightly narrower than the distribution from the source, and shifted to slightly higher mean speed, because the efficiency of the magnetic lens and transverse cooling combination depends on the forward speed. The black curve shows the slowed distribution. Of the ultracold molecules in the beam, approximately 15\% are slowed to below 10~m/s and are within the radial cutoff.  These molecules are about 1000 times colder in the longitudinal dimension than the unslowed distribution, illustrating that both slowing and cooling mechanisms are effective.  The chirped-slowing method results in molecules with a wide range of longitudinal positions but all with very similar forward speeds, as shown in Figure~\ref{Fig:LaserSlowingFigure}(c).  Here, the longitudinal phase-space distribution after slowing indicates that slow molecules are distributed across a region $\approx$0.3~m wide.
Figure~\ref{Fig:LaserSlowingFigure}(d) shows the arrival time and forward speed of molecules at the same detector.  After early times, where unslowed or only partially slowed molecules pass this point, nearly all the arriving molecules have a forward speed between 8 and 10~m/s. 

Scattering of photons from the slowing beam increases the transverse temperature to about 2~mK, at which temperature the rms radius of the beam will grow to 30~mm over 100~ms. If needed, this heating can be reduced by using a weakly converging slowing laser~\cite{Truppe2017}.  Alternatively, the molecules can be re-cooled to the ultracold regime using a single pass of the 2D transverse-cooling light. The beam path can be short in this region because the  beam is so slow.  This combination results in an exceedingly slow and cold molecular beam which can be used directly for a beam-based eEDM measurement.  Alternatively, the molecules can be captured in a magneto-optical trap (MOT) and used for a trap-based measurement. In the following, we discuss both options.

\section{Measuring the eEDM using an ultracold beam}
\label{sec:beamMeasurement}

In this section, we consider an eEDM measurement using the slow, ultracold molecular beam described above.  At transverse temperatures below 100~$\mu$K, a spin-precession time of a few hundred milliseconds is possible before the beam becomes too diffuse to detect efficiently.  Taking $\tau=100$~ms as a conservative spin-precession time and assuming a beam velocity of 10~m/s, the spin-precession region is 1~m long. The size of this region, where we require high electric field and low magnetic field noise, is similar to that of existing beam-based eEDM experiments.   

After laser cooling and slowing, the population is distributed amongst the 12 Zeeman sub-states of $N=1$, whereas the eEDM measurement requires molecules in $N=0$ (see figure \ref{Fig:EnergyLevels}(a)). The population can be transferred efficiently to $|N,F,m\rangle = |0,0,0\rangle$ using the optical pumping scheme described in detail in~\cite{Ho2020}.  The eEDM measurement then proceeds as follows. Lasers drive a Raman transition that prepares the molecules in an equal superposition of $m=+1$ and $m=-1$ within the state $|N,F\rangle = |0,1\rangle$. This state evolves for a time $\tau$, during which the electron's electric dipole moment interacts with the electric field $E_{\rm eff}$, and its magnetic dipole moment interacts with a small applied magnetic field $B$. A second Raman transition projects any part of the state that remains in the original superposition back into $F=0$. Following this sequence, the populations in $F=0$ and $F=1$ are proportional to $\cos^2(\phi_{\rm E} + \phi_{\rm B})$ and $\sin^2(\phi_{\rm E} + \phi_{\rm B})$ respectively, where $\phi_{\rm E} = -d_e E_{\rm eff}\tau/\hbar$ is due to the eEDM interaction and $\phi_{\rm B} = g \mu_{\rm B} B \tau/\hbar$ is due to the Zeeman interaction. Here, $E_{\rm eff}$ is the effective electric field induced by the applied field $E$, and $g$ is the magnetic $g$-factor. Finally, the molecules pass through a detector which measures the population in the two states using the efficient state-selective laser-induced fluorescence detection method described in~\cite{Ho2020}. We assume a detection efficiency of 50\%.

Let us consider the shot-noise limited eEDM sensitivity that can be reached by this laser-cooled molecular beam approach.  Using the estimates given in sections \ref{Sec:Transverse} and \ref{Sec:Slowing} yields $N_{\rm det} = 7 \times 10^{6}$ per shot. The corresponding statistical sensitivity given by equation (\ref{eq:statSensitivity}) is $\sigma_{d_e} = (7 \times 10^{-29} /\sqrt{n_{\rm shot}})\,e$~cm, where $n_{\rm shot}$ is the number of shots in the measurement, and we have assumed $E_{\rm eff} = 17.5$~GV/cm. Taking a repetition rate of 10 shots per second, which is typical for buffer gas sources, and a duty cycle of 0.5 to account for the dead time needed to reverse the electric field, this translates into $\sigma_{d_e} = (1.1 \times 10^{-31} /\sqrt{n_{\rm day}})\,e$~cm, where $n_{\rm day}$ is the number of days of measurement time. 

A uniform high electric field must be generated throughout the spin-precession region, together with a small low-noise magnetic field. The requirements on these fields are rather stringent. For the electric field, our design goal is $E=18$~kV/cm, which gives the assumed $E_{\rm eff} = 17.5$~GV/cm~\cite{Titov1996, Kozlov1997}. For magnetic field control, a four-layer magnetic shield with internal shim coils can reduce the background magnetic field below 1~pT.  An applied uniform magnetic field of 89~pT tunes $\phi_{\rm B}$ to $\pi/4$ where the sensitivity to phase changes is maximised. Magnetic field noise can severely limit the statistical sensitivity if not adequately controlled. The four-layer shield will reduce external magnetic noise to a low enough level, but there is still magnetic Johnson noise from nearby conductors in the experiment, and from the shields themselves~\cite{Rabey2016}. This will be minimised by using alumina electric field plates coated with a thin layer of titanium nitride, and a vacuum chamber constructed from a combination of glass tubing and titanium.  We have modelled this design and find that the Johnson noise at room temperature is equivalent to an eEDM statistical sensitivity of (1$\times 10^{-31}/\sqrt{n_{\rm day}})\,e$~cm, about the same as the shot-noise limited sensitivity.    

Most systematic errors arising in a molecular beam EDM measurement have been considered previously~\cite{Kara2012}. Many can be measured directly by the molecules, often by deliberately exaggerating each imperfection in the experiment and then scaling back to the actual size of that imperfection. The uncertainties in determining systematic shifts of this type scale with the statistical uncertainty of the eEDM measurement, and since most imperfections can be exaggerated enormously these uncertainties are unlikely to be the limiting ones. The motional magnetic field and geometric phase effects are straightforward to control at the $10^{-31}\,e$~cm level. A magnetic field that reverses with $E$ is likely to be the most difficult systematic effect to control. Leakage currents between the electric field plates, or between the plates and ground, can cause this. These currents will be actively monitored during data taking with a resolution better than 1~pA~\cite{Swallows2013}. The measured leakage current correlating with the $E$-reversal must be kept below $10$~pA in order to reduce the systematic uncertainty below $10^{-31}~e$~cm.  Simultaneously, the background magnetic field can be monitored using an array of vapor-cell magnetometers, each having a noise floor of about $50$~fT/$\sqrt{\rm Hz}$ at the low frequency available for reversing $E$~\cite{Abel2020}. Using 20 magnetometers placed along the beamline, the component of magnetic field that correlates with the $E$-reversal can be measured with an uncertainty of 38~aT in one day, which corresponds to an eEDM uncertainty of (1.5$\times 10^{-31}/\sqrt{n_{\rm day}})\,e$~cm.

We see that a measurement using a slow, ultracold beam has the potential to measure the eEDM at a level well beyond the current state of the art. Alternatively, the slow ultracold molecular beam could be loaded into a trap.  A measurement in a trap is an appealing alternative because $\tau$ can be much longer and the experiment can occupy a much smaller volume, which makes it easier to control the electric and magnetic fields at the required levels. We discuss this approach in the next sections. 

\section{Magneto-optical trap}
\label{Sec:MOT}

The first step towards a measurement with trapped molecules is to capture the slow molecular beam in a magneto-optical trap (MOT).  Magneto-optical trapping is considerably more difficult for molecules than for atoms~\cite{Tarbutt2018}. Many laser wavelengths are usually needed to close the cycling transition sufficiently, significantly higher laser power is needed, and the hyperfine structure is often more complicated, requiring several radio-frequency sidebands to be applied to each laser. There are also some more fundamental problems~\cite{Tarbutt2015}. First, the Zeeman splitting of the ground and excited states is not usually as amenable to magneto-optical trapping as in the atomic case. Second, the laser cooling transition has Zeeman sub-levels that are dark to one handedness of circularly-polarized light. As a result, a molecule tends to be optically pumped back and forth between a state that is dark to one beam, then a state that is dark to the other. In the worst case, this results in equal scattering rates from the two counter-propagating beams and no net trapping force. These difficulties have been solved in two ways. In the first, known as an rf MOT, the polarization handedness of each laser beam and the current in the magnetic field coils are synchronously switched at high frequency~\cite{Hummon2013, Norrgard2016}. In the second, known as a dual-frequency MOT~\cite{Tarbutt2015b}, transitions are addressed using two oppositely-polarized frequency components, one red-detuned and the other blue-detuned. Both methods avoid optical pumping into dark states and recover a strong confining force. Using these methods, MOTs of SrF~\cite{Barry2014, McCarron2015, Norrgard2016}, CaF~\cite{Truppe2017b, Williams2017, Anderegg2017} and YO molecules~\cite{Collopy2018} have been developed.

To understand how best to make a MOT for YbF molecules, we use a rate equation model to simulate the MOT numerically~\cite{Tarbutt2015}. The model takes into account all frequency components of the six MOT beams, the magnetic field produced by the coils, all the relevant levels of the molecule, their exact Zeeman shifts, and the transition strengths between all the levels. Comparison with experimental results for SrF and CaF shows that this model gives accurate results for most MOT parameters~\cite{McCarron2015, Norrgard2016, Williams2017}. In YbF, the hyperfine structure of the laser cooling transition adds another challenge to making a MOT. As shown in figure~\ref{Fig:EnergyLevels}(a), there are two almost degenerate hyperfine levels, one with $F=0$ and the other with $F=2$. For a given circular polarization, the force acts in one direction when the molecule is in $F=0$, but in the opposite direction when in $F=2$. Our study of the MOT forces suggests that they are largest when using the four frequency components indicated by the dark green arrows in figure~\ref{Fig:EnergyLevels}(a). Three are red-detuned from their nearest hyperfine component and have the same handedness, while the fourth is blue-detuned and has the opposite handedness. This is the configuration used in our simulations. 

\begin{figure}[t]
	\centering
	\includegraphics[width=\columnwidth]{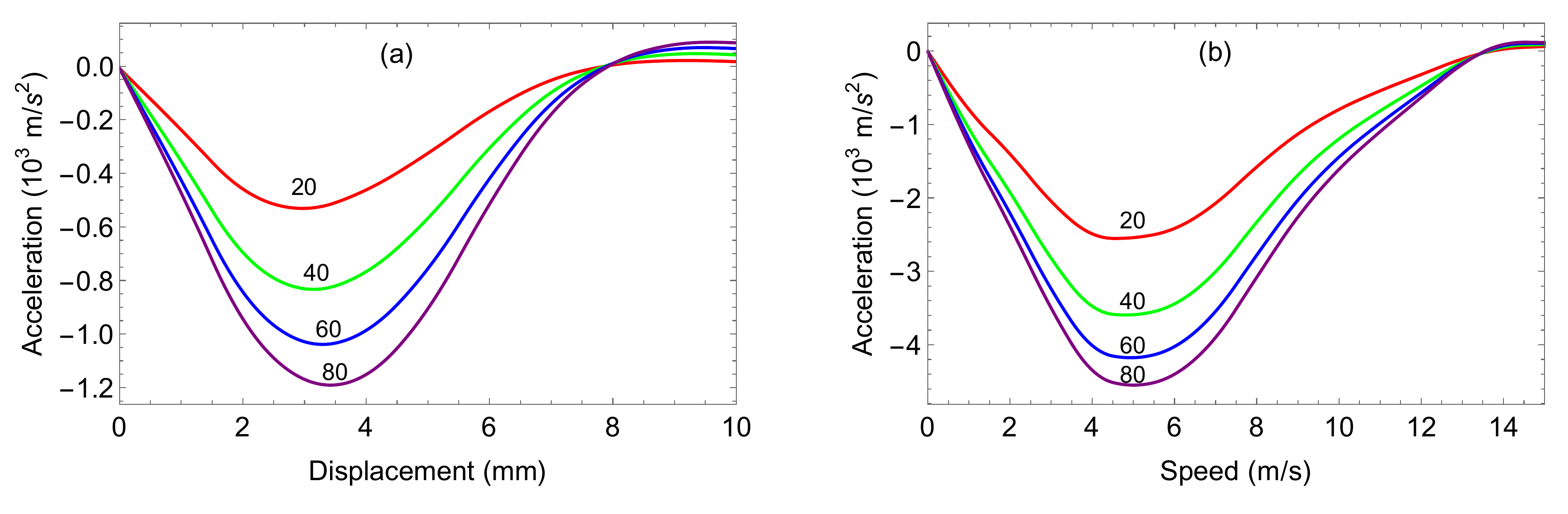}
	\caption{Simulated performance of a magneto-optical trap for YbF molecules. The hyperfine components of the A$^{2}\Pi_{1/2}$--X$^{2}\Sigma^{+}$ (0-0) laser cooling transition are driven using the four near-resonant frequency components shown in figure \ref{Fig:EnergyLevels}(a). Curves show the acceleration versus (a) displacement and (b) speed, for several values of the power per frequency component, in mW. \label{Fig:MOT}}
\end{figure}

Figure~\ref{Fig:MOT}(a) shows the simulated acceleration of a molecule at rest as a function of its axial displacement from the centre of the trap, for a few different values of the power in each frequency component. The MOT beams are Gaussian with $1/e^2$ intensity radii of 12~mm, and the magnetic field gradient is 10~G/cm. There is a confining force for displacements up to $\pm 8$~mm. When using 60~mW per component, the maximum acceleration is $10^{3}$~m/s$^2$ and the axial trap frequency is 100~Hz. With the same trap parameters, figure~\ref{Fig:MOT}(b) gives the acceleration of a molecule at the centre of the trap  versus its speed, showing that there are strong cooling forces for speeds up to 14~m/s. By simulating a beam of molecules entering the MOT region, we calculate a capture velocity of 10.5~m/s. The calculated trap frequency and capture velocity are similar to those measured for CaF and SrF, so we can reasonably expect a MOT of YbF to be similar to the MOTs of molecules already produced.  

In sections \ref{sec:source}, \ref{Sec:Transverse} and \ref{Sec:Slowing} we estimate that $1.5 \times 10^{7}$ molecules per pulse will be made suitable for capture by the MOT. For a CaF MOT, it has been found that the number of molecules actually trapped is about half the number that appear to be within the capture velocity and capture volume~\cite{Williams2017}, and we assume the same fraction for the YbF MOT. If the MOT lifetime is similar to the period over which molecules arrive at the MOT, the fraction accumulated will be further reduced. Figure~\ref{Fig:LaserSlowingFigure}(d) shows that slow molecules arrive at the MOT location, 1.4~m from the buffer gas source, for a period of about 50~ms. We do not currently know the size of residual leaks out of the cooling cycle, but we expect that any leaks can be identified and plugged to achieve a MOT lifetime of at least 50~ms. Thus, we reduce our estimate of the number captured by a further factor of 2, yielding $4\times 10^{6}$ molecules in the MOT. This is about 40 times higher than the current state of the art~\cite{Cheuk2018,Ding2020} with the improvement coming from the powerful combination of magnetic guiding, transverse cooling, and frequency-chirped slowing.

\section{Sub-Doppler cooling}

Once the molecules are captured in the MOT, we will cool them to low temperature using the sub-Doppler cooling methods demonstrated in 3D for CaF, SrF and YO, and in 1D for YbF~\cite{Truppe2017b, McCarron2018, Cheuk2018, Caldwell2019, Ding2020}. The MOT magnetic field is turned off and the molecules are transferred to a blue-detuned optical molasses. In the region where the molasses beams overlap, both the intensity and the polarization of the light field vary over the distance of a wavelength. This causes the ac Stark shift to modulate on the same scale, setting up potential hills for molecules in bright states. Molecules tend to be optically pumped into dark states in places where the intensity is high and the light shift large, and tend to return to bright states in places where the intensity is low and the energies of the dark and bright states are close. As a result, when the light is blue detuned so that the ac Stark shift is positive, molecules climb potential hills and lose energy. This is a form of sub-Doppler cooling, and is able to cool the molecules to very low temperature. Temperatures of about 5~$\mu$K have been observed with CaF and YO, and we expect that YbF will reach a similar temperature. 

\section{Experiment in an optical lattice}

\begin{figure}
    \centering
    \includegraphics[width=0.6\columnwidth]{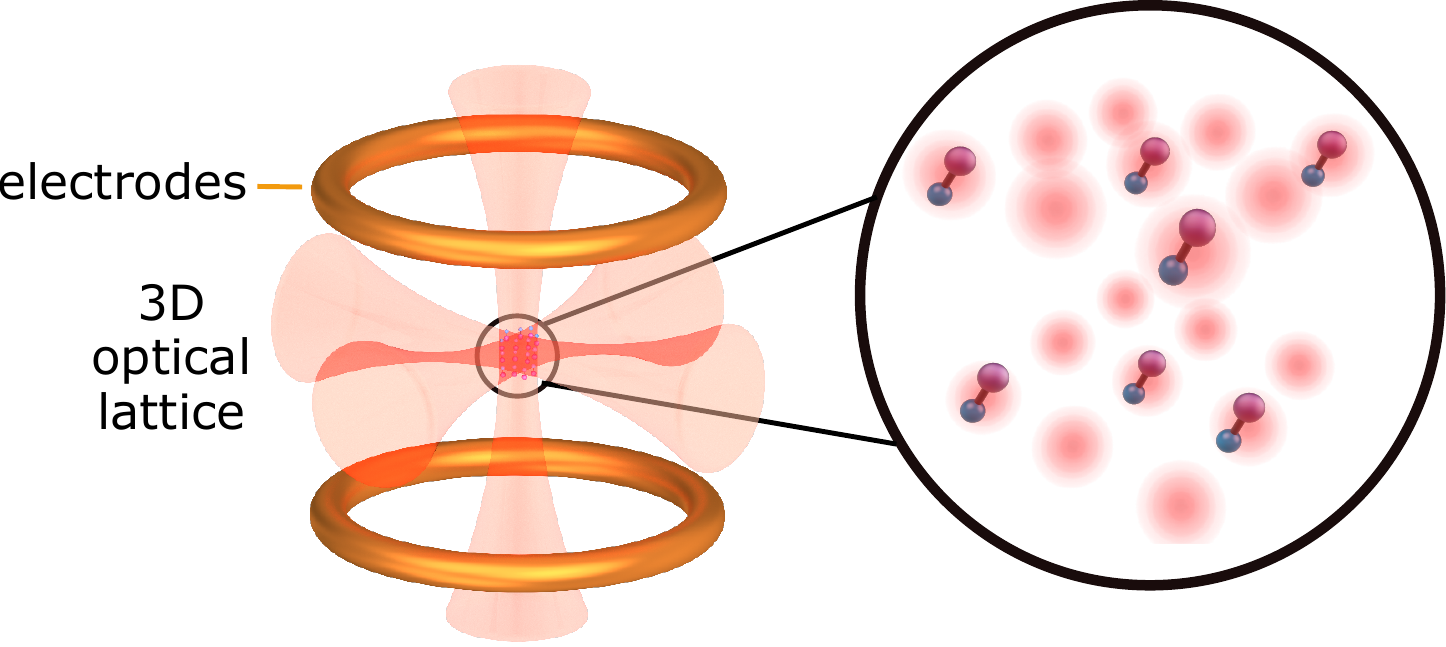}
    \caption{Illustration of a 3D optical lattice of YbF molecules for measuring the eEDM.}
    \label{fig:lattice}
\end{figure}

Let us now consider how an eEDM measurement could be done using these ultracold YbF molecules trapped in an optical lattice. An illustration of such an approach appears in figure~\ref{fig:lattice}.  Here, individual molecules are confined at the anti-nodes of a 3D optical lattice, with the (reversing) electric field used for the eEDM measurement being generated by a pair of ring electrodes.  We propose using a far-detuned lattice with a wavelength of 1064~nm, formed from three orthogonal pairs of orthogonally-polarized counter-propagating laser beams that make a simple cubic lattice. We can estimate the polarizability of the molecule by considering only the $A^{2}\Pi_{1/2}-X^{2}\Sigma^{+}$, $A^{2}\Pi_{3/2}-X^{2}\Sigma^{+}$ and $B^{2}\Sigma^{+}-X^{2}\Sigma^{+}$ transitions with angular frequencies $\omega_{\rm AX,\frac{1}{2}}$, $\omega_{\rm AX,\frac{3}{2}}$ and $\omega_{\rm BX}$ respectively. In this case, when the angular frequency of the laser is $\omega_{\rm L}$, the scalar polarizability is
\begin{equation}
\alpha_0 = \frac{1}{3}(\alpha_{\|}+\alpha_{\perp,\frac{1}{2}}+\alpha_{\perp,\frac{3}{2}})
\end{equation}
where
\begin{eqnarray}
\alpha_{\|} &\approx \left( \frac{1}{\hbar(\omega_{\rm BX}+\omega_{\rm L})} + \frac{1}{\hbar(\omega_{\rm BX}-\omega_{\rm L})} \right) |\mu_{\rm BX}|^2, \\
\alpha_{\perp,\Omega} &\approx \left( \frac{1}{\hbar(\omega_{\rm AX,\Omega}+\omega_{\rm L})} + \frac{1}{\hbar(\omega_{\rm AX,\Omega}-\omega_{\rm L})} \right) |\mu_{\rm AX}|^2. 
\end{eqnarray}
Here, $\mu_{\rm AX}$ is the A--X electronic transition dipole moment in the molecular frame, which has been measured~\cite{Zhuang2011}. We could not find a value for $\mu_{\rm BX}$ in the literature so we take $\mu_{\rm BX} \approx \mu_{\rm AX}$, following the typical pattern of the alkaline-earth monohalides~\cite{Dagdigian1974}. In  this way, we estimate the scalar polarizability at 1064~nm to be $\alpha_0 \approx 1.5 \times 10^{-39}$~J/(V/m)$^2$. This is an underestimate since we have neglected the higher-lying states. Using Gaussian beams, each with waist size 180~$\mu$m and power 25~W, we estimate a lattice trap depth of 39~$\mu$K.  Molecules are loaded into the lattice in the presence of the laser cooling light, and the light shifts are small, so the molecules will have the same temperature in the lattice as they do in free space. They can then be prepared in a single internal state using the coherent control methods demonstrated previously~\cite{Williams2018}.  Loading the lattice requires a relatively high initial density. The highest density of laser-cooled molecules achieved so far is $6 \times 10^8$~cm$^{-3}$, starting from an initial sample of $10^{5}$ molecules~\cite{Cheuk2018}. Using our larger number of molecules, and the rapidly-advancing techniques for compressing and cooling these samples~\cite{Caldwell2019, Ding2020}, a density in the range $10^{11}-10^{12}$~cm$^{-3}$ is a reasonable projection. At this density, $10^{5}-10^{6}$ molecules could feasibly be loaded into the central part of the lattice. Since the density of latice sites is $7\times 10^{12}$~cm$^{-3}$ , the lattice would be sparsely loaded, leaving scope for still larger numbers of molecules.

Once in the lattice, the molecules can be electrically polarized. A highly uniform electric field can be produced using the pair of ring electrodes shown in figure~\ref{fig:lattice}~\cite{Jungerman1984}. This is a convenient geometry because of the excellent optical access it offers. The eEDM can then be measured using the same method as described in section~\ref{sec:beamMeasurement}. A stimulated Raman process spin polarizes the molecules, the spins precess in the presence of the electric field, and the precession angle is read out using a second stimulated Raman process. The sensitivity is proportional to the spin coherence time which we would like to make as long as possible. Vibrational heating due to blackbody radiation limits the coherence time to 4.1~s at 293~K, increasing to 4700~s at 77~K~\cite{Buhmann2008}. The photon scattering rate in such a far-detuned lattice is below 0.1~s$^{-1}$, and the collision rate with background gas atoms is below 0.1~s$^{-1}$ at readily achievable vacuum levels. Thus, it seems feasible to reach spin coherence times of a few seconds at room temperature, and perhaps tens of seconds with modest cooling of the immediate environment. 

An experiment that has $10^{6}$ YbF molecules in the lattice, a spin coherence time of 10~s, and an applied electric field of 18~kV/cm, has a shot-noise limited statistical sensitivity of $2\times 10^{-32}\,e$~cm in one day of measurement. We need to consider how to control noise sources and systematic errors at this level. The most troublesome noise source is likely to be magnetic. With a four-layer shield surrounding the apparatus, the background magnetic field can be reduced to the sub-nT level. Shim coils can be used to reduce the field at the molecules further, and to apply a well controlled bias field. The shot-noise limit corresponds to a magnetic field noise of 2~fT/$\sqrt{{\rm Hz}}$ at the low frequencies needed for switching the electric field; magnetic noise greater than this will compromise the sensitivity of the measurement. Within a multi-layer magnetic shield, the residual magnetic noise tends to be limited by the noise generated by the shields themselves. This is caused by thermal fluctuations of the magnetic domains (magnetization fluctuations) and by thermal fluctuations of the conduction electrons (Johnson noise). With careful design of the shields, noise levels below 1~fT/$\sqrt{{\rm Hz}}$ have been achieved~\cite{Dang2010, Yashchuk2013}. Johnson noise from conductors within the shields is another concern. To mitigate this, the electrodes can be made from alumina coated with a thin layer of titanium nitride. For these electrodes at room temperature, we calculate a Johnson noise of approximately 1$\times$10$^{-32}\,e$~cm in one day, a little below our required noise floor.  Magnetic field gradients result in a spread of the spin precession angle which reduces the sensitivity of the experiment. To ensure that the loss of sensitivity is below 10\%, we require field gradients to be less than 17~pT/cm. With careful degaussing, gradients below 3~pT/cm have been reported for a 2-layer shield without any shim coils~\cite{Altarev2015}. 

Most systematic effects in the lattice will be the same as those discussed in section~\ref{sec:beamMeasurement}. In particular, we note that the magnetic field correlated with electric field reversal needs to be kept below 3~aT to reach an eEDM limit of $10^{-32}\,e$~cm. There are also systematic effects specific to the lattice experiment. One that seems especially difficult to control is the vector light shift. If the lattice light has a component of circular polarization, there will be a vector light shift whose effect is the same as that of a magnetic field. Components of this effective magnetic field perpendicular to the applied electric field have a negligible effect~\cite{Hudson2002}, so we need only consider the parallel component. Taking the $z$-axis along the electric field, and light of polarization $\vec{\epsilon}$ and intensity $I$ at the position of the molecule, the vector Stark shift of the states $|N,F,m_F\rangle=|0,1,m_F\rangle$ is
\begin{equation}
W_1 = \frac{\alpha^{(1)}}{6\varepsilon_0 c} m_{F} \epsilon_{\rm circ} I, 
\end{equation}
where $\epsilon_{\rm circ} = |\epsilon_1|^2 -|\epsilon_{-1}|^2$ is a measure of the circular handedness in the $z$ direction, and $\alpha^{(1)}$ is the vector polarizability given by~\cite{Caldwell2020b}
\begin{equation}
\alpha^{(1)} \approx \frac{1}{2}\left( \frac{\omega_{\rm L}}{\omega_{\rm AX,\frac{3}{2}}}\alpha_{\perp,\frac{3
}{2}} -  \frac{\omega_{\rm L}}{\omega_{\rm AX,\frac{1}{2}}}\alpha_{\perp,\frac{1}{2}}  \right) .
\end{equation}
We estimate $W_1/h \approx -7.1 \times 10^{-6} m_{F} \epsilon_{\rm circ} I$~Hz with $I$ in (W/m$^2$). The lattice beams should be linearly polarized, with orthogonal beams having a small frequency offset to average away any residual interference between them. We note that linear polarization perfection has previously been measured at the $10^{-10}$ level~\cite{Zhu2013}. Being more conservative, let us assume a residual circular handedness of $\epsilon_{\rm circ}=10^{-6}$. In this case, the vector light shift at the antinodes of the lattice is $W_1/h \approx - 42 m_F$~mHz, which is equivalent to a magnetic field of 3.0~pT. This will easily be detected as a phase shift proportional to lattice intensity, and that measurement can be used to minimize the effect and to check for systematic errors. Intensity or polarization fluctuations lead to fluctuations in $W_1$ which can lead to excess noise in the measurement. In the case where $\epsilon_{\rm circ}=10^{-6}$, the eEDM shot-noise limit corresponds to a fractional noise in either $I$ or $\epsilon_{\rm circ}$ of $6 \times 10^{-4}/\sqrt{\rm Hz}$. To prevent the light shift generating a false eEDM at the level of $10^{-32}\,e$~cm, the change in $\epsilon_{\rm circ}$ that correlates with the electric field reversal must be smaller than $10^{-12}$. Similarly, if $\epsilon_{\rm circ}=10^{-6}$, the fractional change in intensity that correlates with the electric field reversal must be below $10^{-6}$. We note that a substantial vector light shifts arises due to large spin-orbit splitting which is a general feature of the heavy molecules needed for eEDM measurements. Thus, the challenging requirements found here for YbF are likely to carry over to any lattice-based eEDM measurement. 

\begin{figure}[t]
	\centering
	\includegraphics[width=0.5\columnwidth]{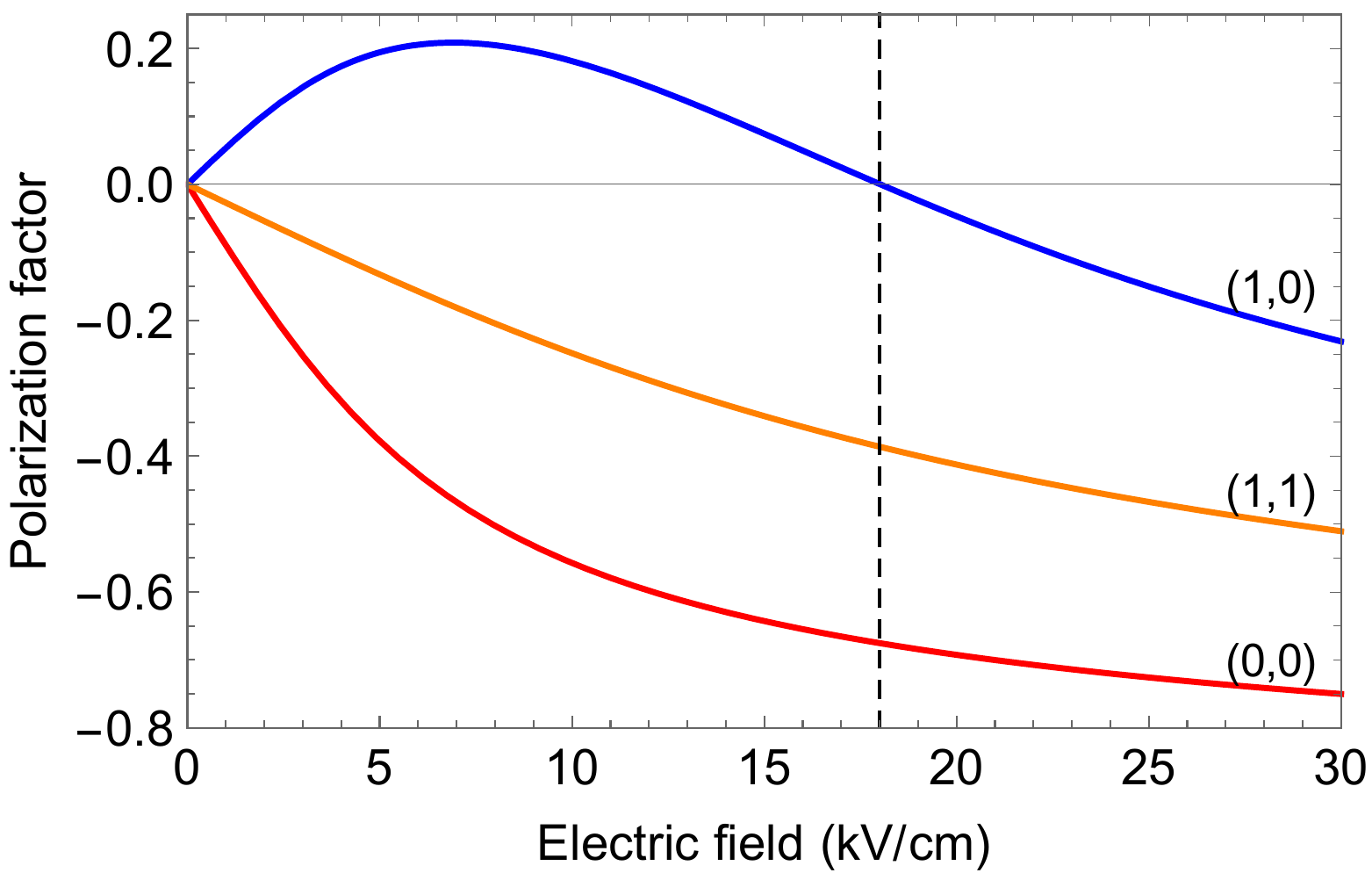}
	\caption{Polarization factor versus electric field for three rotational states of YbF labelled by the quantum numbers $(N, M_N)$.}
	\label{Fig:PolFac}
\end{figure}

The control of systematic errors induced by magnetic fields and light-shifts can be helped enormously by co-magnetometry, which can be done using the molecules themselves. This is possible because the molecule provides states that are insensitive to the eEDM but are fully sensitive to most systematic effects. The magnetic effects are proportional to the spin polarization, whereas the eEDM is proportional to the product of the spin polarization and the electric polarization of the molecule. Figure~\ref{Fig:PolFac} shows the electric polarization factor, $\zeta$, for three rotational states of YbF, as a function of applied electric field. For the eEDM measurement it is natural to use the $|N,m_N\rangle=|0,0\rangle$ state where $\zeta = -0.68$ at 18~kV/cm. At this field, the $|0,0\rangle$ and $|1,0\rangle$ states have similar sensitivity to magnetic fields and light shifts, but the $|1,0\rangle$ state is insensitive to the eEDM because it has  $\zeta = 0$. Furthermore, the $|1,1\rangle$ state provides a pair of $M_F=\pm 1$ levels whose magnetic $g$-factor is 14 times smaller than in $|0,0\rangle$, meaning that it is insensitive to both eEDM and to magnetic fields.  A lattice filled with a mixture of molecules in these various states could therefore provide excellent discrimination of the eEDM signal from systematic effects. 

\section{Conclusions}
Measurements of electric dipole moments are powerful probes of physics beyond the Standard Model. We have shown how ultracold YbF molecules could be used to measure the eEDM with an uncertainty at the $10^{-31}$--$10^{-32}\,e$~cm level. The combination of magnetic focussing followed by two-dimensional transverse cooling and frequency-chirped slowing is a powerful way to produce a slow, highly-collimated molecular beam. Using detailed numerical modelling to optimise the parameters of all three steps, we find that a beam can be produced containing $1.5 \times 10^{7}$ ultracold molecules at speeds below 10~m/s. This beam can be used directly for an eEDM measurement, or the molecules could be loaded into a magneto-optical trap and subsequently into an optical lattice. Our modelling suggests that the capture velocity, spring constant, and damping coefficient of a YbF MOT will be similar to those of other molecular MOTs already produced. Magnetic noise can be a major hindrance to reaching the shot-noise limited sensitivity of the experiment. However, using multi-layer magnetic shields, the noise floor needed to reach a sensitivity of $10^{-32}\,e$~cm in one day has been demonstrated previously. Strategies for reducing Johnson noise to the required level include the use of ceramic electric field plates with thin titanium nitride coatings, glass vacuum chambers, and the selection of titanium wherever metals are unavoidable. Of the many challenging systematic effects to consider, the magnetic field that correlates with the electric field reversal is likely to be the most troublesome. Using currently available vapour cell magnetometers, the effect can be measured with an uncertainty corresponding to $10^{-31}\,e$~cm in one day. This may be the limiting factor in eEDM sensitivity, so improvements in magnetometer sensitivity would make a major contribution to future eEDM measurements. We have considered the vector light shift in a lattice experiment which has a similar effect to a magnetic field. Variations in the intensity of the circularly polarised component produce excess noise, while changes in this component that correlate with electric field reversal produce a false eEDM. The control of these effects leads to demanding requirements for the purity of linear polarization and the intensity stability of the lattice. We have indicated how different rotational states of the molecule can be used to help control systematic errors due to magnetic fields and light shifts. The methods discussed in this paper will also be useful for other tests of fundamental physics. For example, a mid infrared lattice clock could be developed based on the vibrational transition, which could be a sensitive probe of varying fundamental constants. Strongly dipolar molecules in optical lattices are also well suited to studying important problems in many-body quantum physics, metrology beyond the standard quantum limit, and quantum information protocols.

\ack
This work was supported by funding in part from the John Templeton Foundation (grant 61104); the Science and Technology Facilities Council (grants ST/N000242/1, ST/S000011/1); the Sloan Foundation (grant G-2019-12505); the Gordon and Betty Moore Foundation (grant 8864); and the Royal Society (grants URF\textbackslash R1\textbackslash 180578 and RGF\textbackslash  EA\textbackslash 181031).  The opinions expressed in this publication are those of the author(s) and do not necessarily reflect the views of these funding bodies.

\section*{References}
\bibliographystyle{iopart-num}
\bibliography{references}

\end{document}